\documentclass[12pt]{article}

\usepackage[english]{babel}

\usepackage{epsfig}
\usepackage{dcolumn}
\usepackage{amsmath}
\usepackage{changebar}
\usepackage{graphicx}

\newcommand{\ba} {\begin{eqnarray}}
\newcommand{\ea} {\end{eqnarray}}
\usepackage{graphicx}
\usepackage{dcolumn} 
\usepackage{bm}      
\usepackage{umlaut}
\begin{document}
\renewcommand{\figurename}{Fig.}
\renewcommand{\tablename}{Tab.}
\title{Screening masses in thermal and dense medium 
}

\author{ A.~Tawfik\thanks{tawfik@physik.uni-bielefeld.de} \\
 {\small Hiroshima University, 1-7-1 Kagami-yama, Higashi-Hiroshima 
 Japan }  
}

\date{}
\maketitle

\begin{abstract}
Screening masses of different hadronic states are studied in thermal and
 dense medium on lattice.  It has been found that screening masses
 increase with  the temperature. In deconfinement phase, chemical potential
  enhances the screening masses. We use the normalization with
 respect to lowest Matsubara frequency to characterize dissolving of
 hadronic bound states at high temperatures. It has been found that
 different hadronic states have different dissolving temperatures and
 their survivals are considerably improved at finite chemical potentials.
\end{abstract}


Spatial correlation functions are often analyzed on lattice, from which
screening masses can be constructed~\cite{taw1}. At low temperatures, screening
masses are identical with pole masses. The latter can be constructed from
temporal correlation functions. Studying temporal correlation functions
on lattice is limited due to limitation of lattice temporal
dimension. At high temperatures, screening and pole masses have
different dependences. In deconfined phase, screening masses are expected to
approach $l\pi T$, the lowest Matsubara frequency. Therefore, we suggest
to use screening masses normalized to lowest Matsubara frequency to
determine the temperatures, at which hadronic bound states dissolve into
free quarks.  In this limit, screening mass does not depend on {\it
parent} bound states. At very high temperatures, one would expect a
universal line of screening masses from different hadronic states,
$m_D=l \pi T a = l \pi/N_{\tau}$, where $a$ is lattice spacing,
$N_{\tau}$ is the time extention in lattice units. The values of $l$ depend
on hadronic states, $l=2$ for mesons and $l=3$ for baryons. The temporal
masses vanish in limit of free fields. Interactions with thermal
and dense medium drive temporal masses away from free field limit. In
other words, differences between temporal and spatial (screening)
masses measure how close is an interacting system to the case of free
quarks.  \\

Screening masses can be used to characterize different physical
quantities. Their values depend on the properties of the 
medium. Therefore, screening masses may provide a useful tool to investigate
the properties of quark-gluon plasma phase. Values of screening masses
can also quantize   
the response of the medium to weak perturbations, like
hadrons. The response likely appears in form of dynamical Debye screening of 
long-range Coulomb potential. There are different definitions for 
screening masses. They are related to inverse of Debye
screening length. They are the static limits of 
longitudinal vacuum polarization tensor $\prod_{00}(p\rightarrow0)$. \\

In lattice QCD simulations, screening masses can be constructed from
exponential falloff of correlators of two Polyakov loops. %
In this letter, we briefly introduce the results from two flavor lattice
simulations at $am_q=0.05$. The lattice size is
$12^2\times 24 \times 6$. Details on 
numerical simulations can be taken from~\cite{qcdtaro3}.
 
\begin{figure}[thb]
\centerline{
\includegraphics[width=7.cm]{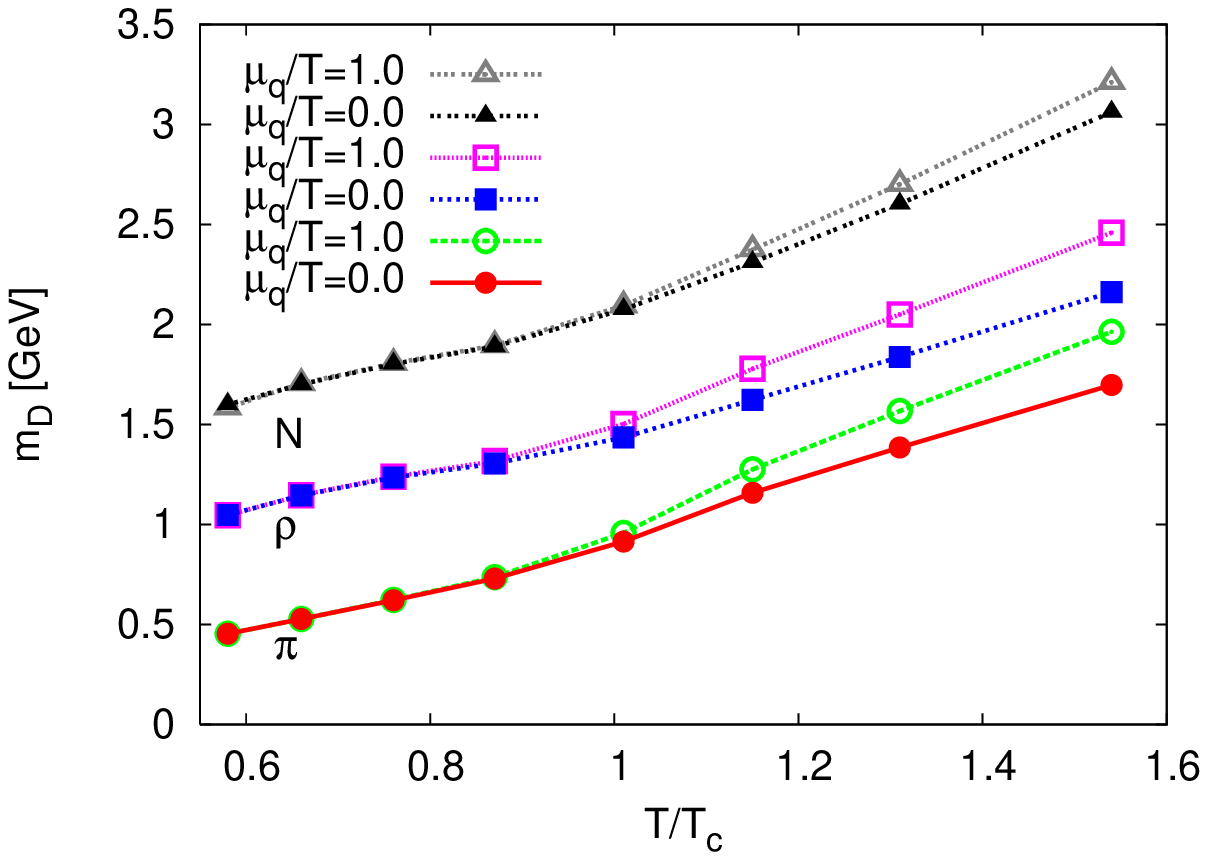}
\includegraphics[width=7.cm]{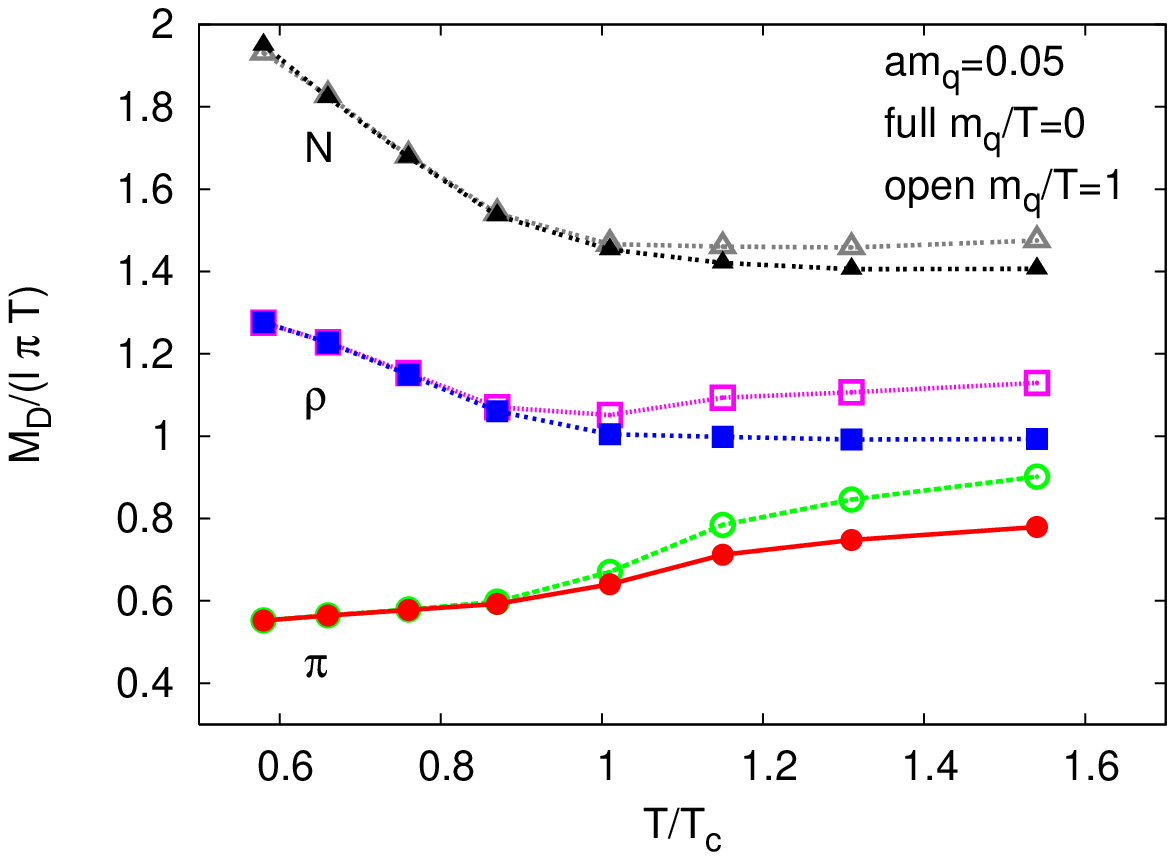}
}
\caption{Left panel: Screening masses, $m_D$  as function of $T/T_c$ for
 three hadronic states at $am_q=0.05$. We set physical units according
 to $T_c=0.225\;$GeV. Right panel: $m_D$ are normalized to lowest Matsubara
 frequency $l\pi T$, where $l=2$ for mesons and $l=3$ for
 baryons. Different hadrons have different dissolving
 temperatures. Survivals are improved at finite chemical potentials. }
 \label{Fig:qcdtaro}  
\end{figure}

To including chemical potentials, we
apply  method of Taylor expansion at fixed temperature, coupling and
bare quark mass.
\ba
\frac{m_D(T,\mu_q)}{T} &=& \left.\frac{m_D(T)}{T}\right|_{\mu_q=0} +
          \left.\left(\frac{\mu_q}{T}\right)\frac{\partial
	  m_D(T)}{\partial\mu_q} 
          \right|_{\mu_q=0} +
	  \left.\left(\frac{\mu_q}{T}\right)^2 \frac{T}{2} 
           \frac{\partial^2 m_D(T)}{\partial\mu_q^2}\right|_{\mu_q=0} +
	   \cdots \nonumber 
\label{Eq:Taylor}
\ea
The coefficients are determined by measuring the hadron
propagators and their derivatives at $\mu=0$ by using standard MC
method~\cite{qcdtaro3,qcdtaro1,qcdtaro2}.  
The results for
degenerate quark masses $am_q=0.05$ are summarized in left panel of 
Fig.~\ref{Fig:qcdtaro}. Error bars are included. They are too small. We
note that the screening mass of each hadron increases with
increasing the temperature. Depending on definitions of screening masses,
we can interpret this result. Response of heat path to the existence of
hadrons raises with temperature. Debye screening length decreases with
temperature and so on. Above $T_c$, this behavior seems to
be stronger. The influence of chemical potential starts to set on above
$T_c$. Below $T_c$, there is no influence. In 
deconfined phase, screening masses at finite $\mu_q$ are larger
than at $\mu_q=0$. Mesons are much sensitive than baryons. It
is worthwhile to note the dependence of screening
masses on hadronic states. We conclude that heat bath
(gluonic system) might be polarized around hadronic states (static charges) and
its parton density seems to be modified accordingly.    \\

Screening masses normalized to lowest Matsubara
frequency, $l\pi T$,  are
depicted in right panel of Fig.~\ref{Fig:qcdtaro}. First, 
we start with results at zero chemical potential. It seems that
$\rho$-meson dissolves much earlier than other hadronic states. In contrast to
$\rho$, $\pi$-meson survives at very high temperature. The nucleon
remains undissolved until $1.5T_c$. Inserting these states in dense
medium (finite chemical potentials) leads to enhancement of their
survivals. For instance, 
$\rho$-meson at $\mu_q/T=1$ can survive at $1.5T_c$. 

Results given in this letter may considerably improve our
understanding of deconfinement, QCD phase structure and critical
phenomenon at high temperatures and densities. 
 
\vspace*{.5cm}

\noindent
{\bf Acknowledgment}\\ 
This work has been financially supported by 
the Japanese Society for the Promotion of Science. 
It is my pleasure to thank Atsushi Nakamura and Irina~Pushkina for useful
discussions. This manuscript is based on talk given at
YITP workshop on ``Hadrons at Finite Density 2006'' 
YITP-W-05-24, Feb. 20-22, 2006.

\end{document}